# Responsible Discovery in Astrobiology:
## Lessons from four controversial claims

**Author's Full Name:** Daliah Bibas, Clément Vidal
**Affiliation:** Center Leo Apostel, Vrije Universiteit Brussel, Belgium

**Short Bios:**
*Dr. Clément Vidal* is a philosopher with a background in logic and cognitive sciences. He is co-founder of the Evo Devo Institute non-profit. In 2022-2023 he was a visiting scholar at UC Berkeley's SETI Research Center. In 2014, he authored *The Beginning and the End: The Meaning of Life in a Cosmological Perspective*. He is eager to tackle big questions, bringing together areas of knowledge such as cosmology, physics, astrobiology, complexity science and evolutionary theory.

*Daliah Bibas* is a doctoral scholar at Vrije Universiteit Brussel, Belgium where she researches the philosophical and ethical implications of discovering non-human intelligence, focusing on extraterrestrial and artificial general intelligences. With a background in astrobiology, space exploration, and science communication, her interdisciplinary interests engage with topics such as the nature of intelligence, the ethical treatment of non-human intelligences, and the long-term future of humanity.

**Abstract:**
This paper examines four case studies of life-detection claims in astrobiology, covering both biosignatures and technosignatures: the 1877 "canals" on Mars, the 1976 Mars Viking landers experiments, the 2020 phosphine detection on Venus, and the 2020 Breakthrough Listen Candidate 1 (BLC1) signal. We analyse the process of discovery for each case, including how they were detected, the media reception, the ensuing scientific debate, the correction processes, and the time it took until an expert consensus was reached. We identify lessons learned while providing scientists, the scientific community, and science communicators with recommendations for approaching future claims of astrobiological discoveries. To avoid potential cognitive biases and mitigate premature conclusions, we stress the need for clear communication of uncertainties, as well as thorough debate and verification processes among the scientific community. These responsible approaches can strengthen the credibility of scientists, cultivate a supportive scientific community, and help astrobiology flourish as a field.

**Introduction**

The field of astrobiology has a history of controversial claims of both biological indicators of extraterrestrial life, termed *biosignatures*, and technological indicators of extraterrestrial intelligence, termed *technosignatures*. There are numerous responsibilities to establish beneficial outcomes for all actors during the process of discovery. Past processes of discovery have been problematic, from scientists lacking high scientific standards to journalists' sensationalising claims. These irresponsible approaches can damage the credibility of scientists, as well as negatively impact the scientific community and the growth of astrobiology as a field. We propose an analysis of four distinct cases of astrobiological claims that include two claims of biosignatures: the 1976 Mars Viking landers case and the 2020 Venus Phosphine case, and two claims of technosignatures: the 1877 Canals on Mars case and the 2020 Breakthrough Listen Candidate 1 (BLC1) case. Our goal is to provide scientists, the scientific community, and science communicators with lessons from the past and recommendations for the future. The discovery of life beyond Earth would arguably be one of the most important scientific discoveries of our time that would not only revolutionize science, but would also radically reshape our philosophical, cultural, and even theological worldviews (see e.g. Dick 2013; Vidal 2015). The stakes, therefore, are extraordinarily high.

**Conceptual background**

We want to start by noting that scientific discovery is rarely ever a single event, but instead, a long process unfolding through different stages including detection, interpretation, scientific debate, correction, and eventual expert consensus. The journey from detection to consensus is often filled with uncertainty: data can be incomplete or ambiguous, methods can be imperfect, and human cognitive biases affect the whole process.

In this paper, we use the responsibility framework of Bower (2025) involving four components: the main actors, the community, the problem, and the appropriate response. The actors we focus on are the scientists making the initial detection, the scientific community debating whether the detection constitutes a proof of extraterrestrial life or intelligence, and the communicators reporting throughout the process. Each of these actors have their own problems and appropriate responses that are not static, but instead evolve with scientific progress, new technological advances, changing societal values, and ideally, lessons learned from previous controversies.

**Case Studies and Lessons Learned**

We now examine four cases that we selected from dozens of controversial astrobiological claims (see Table 1 for a summary). We chose these cases not only because they happened at different historical times, but also because they represent two different kinds of biosignatures: surface and atmospheric; and two different kinds of technosignatures: artefact and radio.

|  | Case Study | | | |
| --- | --- | --- | --- | --- |
| Feature | Canals on Mars (1877) | Viking Landers (1976) | Venus Phosphine (2020) | BLC1 (2020) |
| Signature | Surface Technosignature | In-situ Biosignature | Atmospheric Biosignature | Radio Technosignature |
| Detection | Telescopic observations of linear features on Mars' surface | Metabolic activity in Martian soil | Phosphine gas in Venus' atmosphere | A narrowband radio signal during SETI observations |
| Media Reception | Sensationalism | Confusion | Excitement to Disappointment | Excitement to Disappointment |
| Scientific Debate | Interpretation questioned | Interpretation questioned | Array calibration questioned | Authenticity questioned |
| Correction Processes | Improved observation methods | Ongoing | Ongoing | Detailed radio frequency interference analysis |
| Expert Consensus | Optical illusion | No consensus; Majority believe in non-biological explanation | No consensus; Majority believe in non-biological explanation | Human interference |
| Time until Consensus | ~90 Years | Ongoing debate for ~50 years | Ongoing debate for ~5 years | ~1 Year |
| Ethical Considerations | Misinterpretation and sensationalism without sufficient evidence | Results presented with caution, acknowledging uncertainties | Initial findings presented with caution, acknowledging uncertainties | Initial data leakage from internal channel; further analysis handled with caution and transparency |

**Table 1**: Summary of main features involved in four case studies of astrobiological claims.

### *Canals on Mars (1877)*

One of the earliest astrobiological controversies arose in 1877 after the Italian astronomer Giovanni Schiaparelli sketched a map of Mars based on visual observations. Using a telescope with limited resolution, he reported seeing dark, straight lines across the Martian surface and referred to them as "canali", an Italian word for "channels" that did not explicitly assign them with natural or artificial origin (Schiaparelli 1878). Schiaparelli's detailed descriptions, confirmation by other astronomers, and mistranslation of "canali" into English as "canals" sparked worldwide speculation of Martian waterways (Dick 1996). Although Schiaparelli suggested only that these structures required further study, public fascination grew quickly due to sensational reporting by newspapers and popular science magazines. By the early 20th century, American astronomer Percival Lowell published a hypothesis that solidified the water canal narrative within popular culture and mainstream media. Lowell (1906) argued that an intelligent civilization had carved canals to transport water from the polar ice caps.

Lowell's extensively published theory captivated the public's imagination, helped by sensational headlines, science fiction, and support from high-profile figures like Alexander Graham Bell (1909). The broader scientific community, however, struggled with the lack of reproducible observations: there was enough ambiguity to sustain debate, yet not enough reliable data to establish a firm consensus. Telescopes at the time lacked the resolution necessary to test the existence of these structures and it took decades of scientific and technological advances for astronomers to conclusively dismiss the canal claim. The dispute finally ended when the spacecraft Mariner 4 in 1965, followed by Mariners 6 and 7 in 1969 and Mariner 9 in 1971, captured close-up images of the Martian terrain. Rather than rows of neatly laid-out canals, the pictures revealed countless craters and other geological features, none of which formed the straight networks that had once spurred belief in an advanced Martian civilization (Sagan and Fox 1975).

This historical example demonstrates how untested early-stage observations can quickly become part of public belief long before they are properly verified. Lowell's widely publicized claim of canals on Mars framed his speculative interpretations as near-certainty. This sensationalist overamplification turned a blurry observation into a cultural myth that stuck for decades. This was not just careless from Lowell's part; it was also a failure of responsibility from the endorsing scientists at the time.

### *Viking Landers (1976)*

NASA's Viking 1 and 2 landers reached the Martian surface in 1976 and performed the first direct life-detection experiments on another world. Each lander carried out three biology experiments: the Labeled Release (LR), which measured gases released by possible microbial metabolism; the Gas Exchange (GEX), which tracked changes in gas composition from soil samples; and the Pyrolytic Release (PR), which tested for photosynthesis-like carbon fixation. In addition, each lander carried a Gas Chromatograph-Mass Spectrometer (GCMS), a chemical analysis instrument, to analyse soil chemistry and detect organic molecules (Klein et al. 1976). The LR experiment, designed by Gilbert Levin and Patricia Straat (1977), exposed Martian soil to a nutrient solution tagged with radioactive carbon-14 and measured any resulting gas. When the soil quickly released carbon dioxide, and both Viking landers thousands of kilometres apart showed the same results, the immediate interpretation was that it may be due to microbial metabolism. However, the GCMS experiment failed to detect organic molecules in the same soil, and the GEX and PR results were interpreted as either ambiguous or negative (Klein et al. 1976).

The conflicting results led NASA to stress in press conferences and in official reports that no definitive evidence of life was found (French 1977; Klein 1977). Media outlets celebrated potential signs of life, while also emphasizing the multitude of unanswered questions. Decades of debates followed. Many scientists argued that the absence of organic compounds and the presence of strong oxidizers in Martian soil made a non-biological explanation more likely (e.g., Mazur et al. 1978; Klein 1999). Meanwhile, others published papers defending the microbial hypothesis (e.g., Aksyonov 1979; Levin and Straat 1981), with many pointing out that perchlorates (highly reactive chlorine-based salts discovered on Mars decades later) could destroy organics during the GCMS heating, therefore explaining the negative results (e.g., Houtkooper and Schulze-Makuch 2010; Navarro-González et al. 2010). Almost half a century later, the majority of the scientific community leans towards an abiotic interpretation, yet a minority still sees the LR results as indicative of extant life on Mars (e.g., Bianciardi et al. 2012; Schulze-Makuch 2024).

Similar to how higher quality observations of Mars disproved Lowell's canals claim, advances in Martian soil analysis from later missions to Mars, such as Curiosity's 2018 discovery of complex organic molecules (Eigenbrode et al. 2018; Freissinet et al. 2025), is gradually changing the views of some scientists regarding the interpretation of Viking's data. However, scientists might be subject to an *anchoring bias*, where they rely too heavily on the first initial conclusions acquired, explaining this slow change of attitudes. Hence, the scientific community might avoid exploring such alternative interpretations with due diligence, whether consciously or not. It is certainly more difficult to navigate the re-opening of past experiments than to discover unambiguous evidence from a new observation or experiment.

### *Venus Phosphine (2020)*

In September 2020, a team of researchers led by astronomer Jane Greaves reported the detection of phosphine gas in Venus' atmosphere (Greaves et al. 2020). Although phosphine is considered a biosignature because it is almost exclusively produced by biological processes on Earth (Sousa-Silva et al. 2020), the authors cautioned that it could also be produced through unknown abiotic chemical processes.

Initially, major news outlets enthusiastically ran the story, but as two independent teams reanalysed both the James Clerk Maxwell Telescope (JCMT) and the Atacama Large Millimeter Array (ALMA) data, issues were identified with ALMA's baseline calibration (Snellen et al. 2020; Villanueva et al. 2021). The media quickly picked up on the controversy and reported that the original claim had ultimately been refuted (e.g., Siegel 2020). Within weeks, the original team refined their analysis using the corrected ALMA data release and clarified that the signal was weaker than initially reported, but that phosphine was still detected (Greaves et al. 2021). Other researchers argued that the signal might have been misinterpreted, possibly resulting from sulphur dioxide contamination or instrumental noise (Thompson 2020; Lincowski et al. 2021). While some scientists remain cautiously optimistic, the growing consensus has shifted toward scepticism, with many concluding that additional evidence is needed before the phosphine detection is generally accepted or can be regarded as a biosignature for Venus (for a review see Clements 2022).

The phosphine detection controversy exemplifies a maturing astrobiology community that is learning to handle ambiguity with greater care. Since the impact of finding life on Venus would be very important scientifically, but also socially and psychologically (Vidal and Bibas 2024), it is certainly desirable that the scientific community keeps being involved, as happened for example with the independent teams who discovered calibration problems in the ALMA data, as well as the response from the initial researchers who showed responsibility by quickly correcting their statements.

### *Breakthrough Listen Candidate 1 (2020)*

Breakthrough Listen is an observational program that searches for extraterrestrial signals. During observations of Proxima Centauri in 2019, an unusual narrowband radio signal was detected and later termed Breakthrough Listen Candidate 1 or BLC1 (Smith et al. 2021). The signal was recorded at a frequency range consistent with artificial and celestial origins and was thought to have been detected in isolation from known human-made sources. In 2020, before the research team had completed their analysis, news of the signal leaked from internal communication channels, which threw BLC1 into the public spotlight (see Sample 2020). Media outlets quickly fuelled widespread speculation with clickbait headlines about the potential discovery of an advanced alien civilization. Although the team behind the detection maintained caution while emphasizing that further analysis was needed to rule out human interference, the narrative had already taken shape in public discourse. Over the next year, the team continued to study the signal using a comprehensive radio frequency interference analysis to investigate whether BLC1 originated from Earth rather than from an extraterrestrial source. In 2021, the team published their first papers about BLC1 which concluded that the signal was the result of human interference (Sheikh et al. 2021; Smith et al. 2021).

The BLC1 case is a prime example of how leaks of unverified data can strain scientific integrity and public trust. Although the anonymous leaker's intentions and motivations are not known, their actions put pressure on the Breakthrough Listen team to resolve the issue and provide the public with a conclusive verdict. Even with this distraction, the team responded responsibly by continuing their analysis with caution, upholding high academic standards, and publicly lowering the expectations that it would be a technosignature, before concluding that the signal was of terrestrial origin.

## Responsibilities for Future Claims

The four cases discussed highlight how different forms of responsibility, or lack thereof, directly shaped the trajectory of each claim. Lessons from each case demonstrate that responsible discovery in astrobiology is not just an abstract ideal, but a practice that unfolds across scientists, the scientific community, and science communicators. This practice involves upholding high academic standards, transparently communicating uncertainties, and recognizing that discovery requires patience while claims undergo methodical testing, debate, and peer review.

The nemesis of good scientific practice is represented by the many pseudoscientific UFO claims. They often follow little-to-no scientific standards, assert absolute certainty, are largely overexaggerated for media attention, and fuel conspiracy theories. A common feature of UFO reports is their ambiguity and transient nature, which leaves a lot of room for interpretation and speculation. By contrast, scientific statements strive to be reproducible and as unambiguous as possible, to minimize uncertainty.

### *Scientists*

Scientists should avoid overinterpretation of ambiguous data and proactively engage with critics from the scientific community to continually gather more data and improve existing models. For instance, astronomer Jane Greaves and her team exemplified this by revising their analysis of phosphine on Venus when calibration errors were identified by other researchers. Furthermore, when making a claim, researchers should always provide a predictive model. If a claim like canals on Mars was made today, it would certainly be labelled as pseudoscience by the scientific community. Although Lowell argued that water was being distributed on the planet from the poles, this theory was not giving rise to new observable predictions. These considerations show that the standards of evidence in science have dramatically increased since the 1870s. Science works best when the uncertainties are well acknowledged and spelled out. Scientists must emphasize that any visual clues (like apparent canals on Mars) or early hunches are just rudimentary and provisional starting points, not conclusions determined from detailed predictive models.

Another key responsibility is to protect investigations from reckless and premature disclosures, something that the Breakthrough Listen team, unfortunately, struggled with. When sensitive data slips out before a thorough analysis is completed, it can fuel speculation, hype, and even conspiracy theories, which ultimately distract the public from the actual findings published later on.

### *Scientific Community*

The first reaction of the scientific community is often to criticize the claims with the aim of deconstructing and debunking, but it is too rarely followed up with the more difficult task of being constructive, for example by proposing new tests and observations.

Resistance against a new idea happens not only between two individual scientists, but also between subcommunities. In astronomy, getting observation time on telescopes is extremely competitive, so it may be hard to plan follow-up observations if the time-allocating committee is sceptical or against the new idea. This can potentially stall any progress as using alternative telescopes would not provide the same baseline. Time-allocation committees need to be acutely aware of cognitive biases that may impede the overall progress of science. For example, since the discovery of extraterrestrial life or intelligence would be paradigm-shifting, a *conservatism bias* is likely, making such riskier or unconventional proposals undervalued because they may not fit within the existing scientific establishment.

### *Science Communicators*

Science communicators, including journalists, science writers, or public information officers, should ensure that new hypotheses are neither prematurely dismissed nor overshadowed by more sensational lines of inquiry. They could inadvertently trigger a strong *anchoring bias* towards their story, especially when they break the news.

The temptation to make sensationalist pieces and choose clickbait titles should be resisted by science communicators. This increase of sensationalist language is not limited to online media, but is a general issue since the 2010s, where language, even in books, gets more and more polarized and cognitively distorted (Bollen et al. 2021). Even when official statements downplay a discovery, media outlets may prioritize sensational angles due to economic pressure, audience expectations, or lack of scientific nuance.

Science communicators could trigger a sense of mystery by spelling out the scientific uncertainties, instead of polarizing hypotheses, and they may also try to portray the scientific debates as suspenseful, instead of taking a side and claiming a definitive resolution. Some space agencies and journals now use embargo systems and coordinated press releases, but there is no universal protocol for life detection claims, making consistency difficult to enforce.

## Conclusion

Responsibilities are involved throughout the entire process of discovery. In the four case studies we analysed, we identified irresponsible actions at various stages. We saw that with the BLC1 case, the unexpected data leak was problematic. For the Viking missions and the detection of phosphine on Venus, the core responsibilities involved are still relevant today and represent typical debates at the heart of the scientific process. For the canals on Mars claim, the most problematic part was the sensationalism and overamplified dissemination. To strengthen public trust in astrobiology, it is imperative that scientists, the scientific community, and science communicators continue building a culture where responsibility is not an afterthought, but embedded at every stage of the process.

The two technosignatures we have examined have been unambiguously refuted, while there remain uncertainties with the possibility of biosignatures on Mars and Venus. This dichotomy suggests that technosignatures may indeed be less ambiguous than biosignatures (Wright et al. 2022; Smith and Mathis 2023). Yet, the discovery of extraterrestrial life or intelligence will likely not arise from one flashy moment, but from a slow, methodical process (Dick 1996).

This also implies that the scientific community should remain open to re-assess past cases, where evidence might be creatively sifted. Even if no conclusive evidence is quickly found, existing hypotheses of positive detections of life beyond Earth may eventually turn out to be pre-discoveries of new astronomical phenomena, as was the case for the discovery of pulsars (rapidly rotating neutron stars that emit regular radio pulses) which were first misinterpreted as possible alien signals (Penny 2013). In other words, pre-discoveries may stay unnoticed in archived observations until brought into a new theoretical light, possibly even generations later.

Taken together, these case studies show that responsibility is multi-layered: scientists must communicate uncertainty and avoid overinterpretation; the scientific community must guard against cognitive biases while enabling follow-up; and communicators must represent ambiguity and resist sensationalism. Embedding these responsible practices at every stage of the discovery process holds the potential to strengthen the credibility of scientists, cultivate a productive scientific community, and help astrobiology flourish as a field.

**Acknowledgements**
We thank David L. Clements and Dirk Schulze-Makuch for sharing their insights on the evolution of the debates of Venus and Mars biosignature detections.

**References**
Aksyonov, S. I. 1979. "Some Comments on Interpretations of Viking Biological Experiments." *Origins of Life* 9 (3): 251–56. https://doi.org/10.1007/BF00932499.
Bell, Alexander Graham. 1909. "Letter from Alexander Graham Bell to Mabel Hubbard Bell, November 29, 1909." November 29. MSS51268: Folder: Mabel Hubbard Bell, Family Correspondence, Alexander Graham Bell, November-December 1909. Manuscript Division. https://www.loc.gov/resource/magbell.04300515/.
Bianciardi, Giorgio, Joseph D. Miller, Patricia Ann Straat, and Gilbert V. Levin. 2012. "Complexity Analysis of the Viking Labeled Release Experiments." *International Journal of Aeronautical and Space Sciences* 13 (1): 14–26. https://doi.org/10.5139/IJASS.2012.13.1.14.
Bollen, Johan, Marijn ten Thij, Fritz Breithaupt, et al. 2021. "Historical Language Records Reveal a Surge of Cognitive Distortions in Recent Decades." *Proceedings of the National Academy of Sciences* 118 (30): e2102061118. https://doi.org/10.1073/pnas.2102061118.
Bower, Adam. 2025. "Responsibility in Outer Space: An IR Perspective." In *Responsibility in Space*. Routledge.
Clements, David L. 2022. "Venus, Phosphine and the Possibility of Life." *Contemporary Physics* 63 (3): 180–99. https://doi.org/10.1080/00107514.2023.2184932.
Dick, Steven J. 1996. *The Biological Universe: The Twentieth Century Extraterrestrial Life Debate and the Limits of Science*. Cambridge University Press.
Dick, Steven J. 2013. "The Societal Impact of Extraterrestrial Life: The Relevance of History and the Social Sciences." In *Astrobiology, History, and Society: Life Beyond Earth and the Impact of Discovery*, edited by Douglas A. Vakoch. Springer. https://doi.org/10.1007/978-3-642-35983-5_12.
Eigenbrode, Jennifer L., Roger E. Summons, Andrew Steele, et al. 2018. "Organic Matter Preserved in 3-Billion-Year-Old Mudstones at Gale Crater, Mars." *Science* 360 (6393): 1096–101. https://doi.org/10.1126/science.aas9185.
Freissinet, Caroline, Daniel P. Glavin, P. Douglas Archer, et al. 2025. "Long-Chain Alkanes Preserved in a Martian Mudstone." *Proceedings of the National Academy of Sciences* 122 (13): e2420580122. https://doi.org/10.1073/pnas.2420580122.
French, Bevan M. 1977. *Mars: The Viking Discoveries*. NASA Educational Publication (EP) NASA-EP-146. NASA. https://ntrs.nasa.gov/citations/19800009678.
Greaves, Jane S., Anita M. S. Richards, William Bains, et al. 2020. "Phosphine Gas in the Cloud Decks of Venus." *Nature Astronomy*, 1–10. https://doi.org/10.1038/s41550-020-1174-4.

Greaves, Jane S., Anita M. S. Richards, William Bains, et al. 2021. "Reply to: No Evidence of Phosphine in the Atmosphere of Venus from Independent Analyses." *Nature Astronomy* 5 (7): 636–39. https://doi.org/10.1038/s41550-021-01424-x.

Houtkooper, J. M., and D. Schulze-Makuch. 2010. "Xerophiles on Mars: Possible Evolutionary Strategies Using Hydrogen Peroxide and Perchlorates." *Evolution and Life: Surviving Catastrophes and Extremes on Earth and Beyond* 1538 (April): 5382. https://ui.adsabs.harvard.edu/abs/2010LPICo1538.5382H.

Klein, Harold P. 1977. "The Viking Biological Investigation: General Aspects." *Journal of Geophysical Research* 82 (28): 4677–80. https://doi.org/10.1029/JS082i028p04677.

Klein, Harold P. 1999. "Did Viking Discover Life on Mars?" *Origins of Life and Evolution of the Biosphere* 29 (6): 625–31. https://doi.org/10.1023/A:1006514327249.

Klein, Harold P., Joshua Lederberg, Alexander Rich, Norman H. Horowitz, Vance I. Oyama, and Gilbert V. Levin. 1976. "The Viking Mission Search for Life on Mars." *Nature* 262 (5563): 24–27. https://doi.org/10.1038/262024a0.

Levin, Gilbert V., and Patricia A. Straat. 1981. "A Search for a Nonbiological Explanation of the Viking Labeled Release Life Detection Experiment." *Icarus* 45 (2): 494–516. https://doi.org/10.1016/0019-1035(81)90048-8.

Levin, Gilbert V., and Patricia Ann Straat. 1977. "Life on Mars? The Viking Labeled Release Experiment." *Biosystems* 9 (2–3): 165–74. https://doi.org/10.1016/0303-2647(77)90026-0.

Lincowski, Andrew P., Victoria S. Meadows, David Crisp, et al. 2021. "Claimed Detection of PH3 in the Clouds of Venus Is Consistent with Mesospheric SO2." *The Astrophysical Journal Letters* 908 (2): L44. https://doi.org/10.3847/2041-8213/abde47.

Lowell, Percival. 1906. *Mars and Its Canals*. Macmillan.

Mazur, Peter, Elso S. Barghoorn, Harlyn O. Halvorson, Thomas H. Jukes, Isaac R. Kaplan, and Lynn Margulis. 1978. "Biological Implications of the Viking Mission to Mars." *Space Science Reviews* 22 (1): 3–34. https://doi.org/10.1007/BF00215812.

Navarro-González, Rafael, Edgar Vargas, José de la Rosa, Alejandro C. Raga, and Christopher P. McKay. 2010. "Reanalysis of the Viking Results Suggests Perchlorate and Organics at Midlatitudes on Mars." *Journal of Geophysical Research: Planets* 115 (E12). https://doi.org/10.1029/2010JE003599.

Penny, Alan John. 2013. "The SETI Episode in the 1967 Discovery of Pulsars." *European Physical Journal H* 38: 535–47. https://doi.org/10.1140/epjh/e2012-30052-6.

Sagan, Carl, and Paul Fox. 1975. "The Canals of Mars: An Assessment after Mariner 9." *Icarus* 25 (4): 602–12. https://doi.org/10.1016/0019-1035(75)90042-1.

Sample, Ian. 2020. "Scientists Looking for Aliens Investigate Radio Beam 'from Nearby Star.'" Science. *The Guardian*, December 18. https://www.theguardian.com/science/2020/dec/18/scientists-looking-for-aliens-investigate-radio-beam-from-nearby-star.

Schiaparelli, Giovanni Virginio. 1878. *Osservazioni astronomiche e fisiche sull'asse di rotazione e sulla topografia del pianeta Marte: fatte nella Reale Specola di Brera in Milano coll'equatoriale di Merz durante l'opposizione del 1877.* Salviucci.

Schulze-Makuch, Dirk. 2024. "We May Be Looking for Martian Life in the Wrong Place." *Nature Astronomy* 8 (10): 1208–10. https://doi.org/10.1038/s41550-024-02381-x.

Sheikh, Sofia Z., Shane Smith, Danny C. Price, et al. 2021. "Analysis of the Breakthrough Listen Signal of Interest Blc1 with a Technosignature Verification Framework." *Nature Astronomy* 5 (11): 1153–62. https://doi.org/10.1038/s41550-021-01508-8.

Siegel, Ethan. 2020. "Venus Is Dead! New Analysis Shows Phosphine, A Possible Biosignature, Is Absent." Forbes, October 22. https://www.forbes.com/sites/startswithabang/2020/10/22/venus-is-dead-new-analysis-shows-phosphine-a-possible-biosignature-is-absent/.

Smith, Harrison B., and Cole Mathis. 2023. "Life Detection in a Universe of False Positives." *BioEssays* 45 (12): 2300050. https://doi.org/10.1002/bies.202300050.

Smith, Shane, Danny C. Price, Sofia Z. Sheikh, et al. 2021. "A Radio Technosignature Search towards Proxima Centauri Resulting in a Signal of Interest." *Nature Astronomy*, 1–5. https://doi.org/10.1038/s41550-021-01479-w.

Snellen, I. A. G., L. Guzman-Ramirez, M. R. Hogerheijde, A. P. S. Hygate, and F. F. S. van der Tak. 2020. "Re-Analysis of the 267-GHz ALMA Observations of Venus: No Statistically Significant Detection of Phosphine." *Astronomy & Astrophysics* 644: L2. https://doi.org/10.1051/0004-6361/202039717.

Sousa-Silva, Clara, Sara Seager, Sukrit Ranjan, et al. 2020. "Phosphine as a Biosignature Gas in Exoplanet Atmospheres." *Astrobiology* 20 (2): 235–68. https://doi.org/10.1089/ast.2018.1954.

Thompson, M. A. 2020. "The Statistical Reliability of 267 GHz JCMT Observations of Venus: No Significant Evidence for Phosphine Absorption." *Monthly Notices of the Royal Astronomical Society: Letters* 501 (1): L18–22. https://doi.org/10.1093/mnrasl/slaa187.

Vidal, Clément. 2015. "A Multidimensional Impact Model for the Discovery of Extraterrestrial Life." In *The Impact of Discovering Life beyond Earth*. Cambridge University Press. https://zenodo.org/record/1286707.

Vidal, Clément, and Daliah Bibas. 2024. "Life in the Clouds of Venus: The Phosphine Debate and Its Scientific, Social, and Psychological Impact." In *IAU Symposium IAUS 387*, edited by H. Landt, M. Dominik, and C. Oliver, To appear. Cambridge University Press. https://doi.org/10.5281/zenodo.13301306.

Villanueva, G. L., M. Cordiner, P. G. J. Irwin, et al. 2021. "No Evidence of Phosphine in the Atmosphere of Venus from Independent Analyses." *Nature Astronomy* 5 (7): 631–35. https://doi.org/10.1038/s41550-021-01422-z.

Wright, Jason T., Jacob Haqq-Misra, Adam Frank, Ravi Kopparapu, Manasvi Lingam, and Sofia Z. Sheikh. 2022. "The Case for Technosignatures: Why They May Be Abundant, Long-Lived, Highly Detectable, and Unambiguous." *The Astrophysical Journal Letters* 927 (2): L30. https://doi.org/10.3847/2041-8213/ac5824.